\documentclass[11pt]{article}
\usepackage[a4paper,margin=25mm]{geometry}
\usepackage[T1]{fontenc}
\usepackage[utf8]{inputenc}
\usepackage{lmodern}
\usepackage{microtype}
\usepackage{amsmath,amssymb}
\usepackage{graphicx}
\usepackage{booktabs}

\usepackage[numbers,sort&compress]{natbib}
\usepackage[colorlinks=true,linkcolor=black,citecolor=black,urlcolor=blue]{hyperref}
\usepackage{caption}

\graphicspath{{./}}
\newcommand{\code}[1]{\texttt{#1}}
\newcommand{\Rsq}{R^{2}}

\title{Reading a neuron's tuning from its correlations with other neurons}
\author{Francesco L\"assig\thanks{Independent researcher.}}
\date{\today}

\begin{document}
\maketitle

\begin{abstract}
What a neuron represents is usually established by relating its activity to a stimulus. We ask whether the same facts are contained in the correlation matrix of a recorded population, which keeps the pairwise relations among the neurons and discards the stimulus. A transformer that sees only the correlation matrix of 512 sampled neurons, with no labels and no reference population in its input, decodes the preferred orientation of mouse visual cortex neurons from in-vivo MICrONS responses to a mean error of $25^\circ$ (uninformed: $45^\circ$; best constant guess: $41^\circ$; $28^\circ$ when the error is averaged over eight orientation bins rather than over neurons) and their receptive-field position at $\Rsq = 0.23$; on the digital-twin substrate, $20^\circ$ ($24.5^\circ$ balanced) and 0.48. Asked whether a neuron prefers horizontal or vertical, it is right for 79\% of neurons in vivo and 87\% on the twin, against 56\% for the majority answer; it does not resolve the obliques. The ordering is recoverable because the relations between orientation classes are close to circulant, which fixes the class structure up to rotation and reflection; the rotation of the frame is recoverable only because of a departure from that symmetry. The departure survives equal representation of every orientation and disjoint training and test stimuli, and it is an anisotropy of the signal correlations: in V1 the neurons preferring the dominant orientation have the most similar responses to one another, and a model population with only that property recovers the frame. Across 33 mice of the Allen Brain Observatory, with relations from drifting gratings, orientation transfers to unseen animals only when a sampled population mixes animals; populations drawn from one animal are not decoded measurably below the best constant, a limit of the diversity of populations one animal provides: its populations of 128 neurons overlap in a large fraction of their neurons, whereas single-scan populations in MICrONS, drawn from scans of about 1,000 neurons, reach $31^\circ$ to $32^\circ$, $9^\circ$ to $10^\circ$ below the MICrONS best constant of $41.3^\circ$. For receptive fields, what crosses animals is where each animal's imaged patch looks on the screen, not the layout of neurons within it.
\end{abstract}

\section{Introduction}

A neuron in visual cortex can be characterised by what it responds to: vertical gratings, or the upper left of the screen. This is conventionally established with the stimulus. We ask whether the same facts are contained in a description that mentions no stimulus. For a population that watched the same movie, one can compute the correlation of every pair of neurons and then discard the movie. Does the resulting matrix determine which neuron prefers which orientation, and where each neuron's receptive field lies?

The idea that content can be read from relations is old. Shepard and Chipman proposed that a representation need not resemble what it stands for as long as the similarities among representations mirror the similarities among the things represented \citep{shepard1970}, and representational similarity analysis turned this into a method that compares brains and models through their similarity structures rather than their coordinates \citep{kriegeskorte2008}. Structuralist accounts of experience claim that what a state represents, or how it feels, is fixed by its position in a space of relations \citep{lyre2022}. \citet{framework2025} formulates the corresponding question for neural networks and gives the decoding framework we use here.

Two lines of work come close. \citet{tsodyks1999} showed that the spike-triggered average of spontaneous population activity reproduces the map of a neuron's preferred orientation, with no stimulus present; the tuning was read from that pattern by comparison with the evoked map, a stimulus-labelled reference. Unsupervised manifold methods recover a neuron's tuning from the structure of population activity alone, without measuring the variable it encodes \citep{rubin2019}, but the recovered coordinate is defined only up to the symmetries of the latent space, and it is related to the measured variable by aligning to it afterwards. In both cases the relations locate a neuron within the structure of the encoded variable, and something outside the relations, the evoked map or the measured behaviour, says which value that location corresponds to. We ask whether the relations among neurons fix the content of each neuron when the stimulus is discarded and there is no labelled neuron, no map and no behaviour to compare with. We also ask whether that content is readable across thousands of individually identified neurons and across animals. The nearest machine-learning work reads properties of a neuron from population activity: NuCLR learns a neuron's identity from its population context with a permutation-equivariant transformer and transfers cell type and brain region to unseen animals \citep{arora2025}. It sees the activity itself. Our decoder sees only the correlation matrix, so it answers what the relations alone determine, and its target is stimulus tuning rather than cell class. This relational setting is common: recordings under natural behaviour have no controlled stimulus, and neurons from different sessions or animals share no stimulus-aligned coordinates, but they do share the structure of their correlations.

\begin{figure}[tbp]
  \centering
  \includegraphics[width=\textwidth]{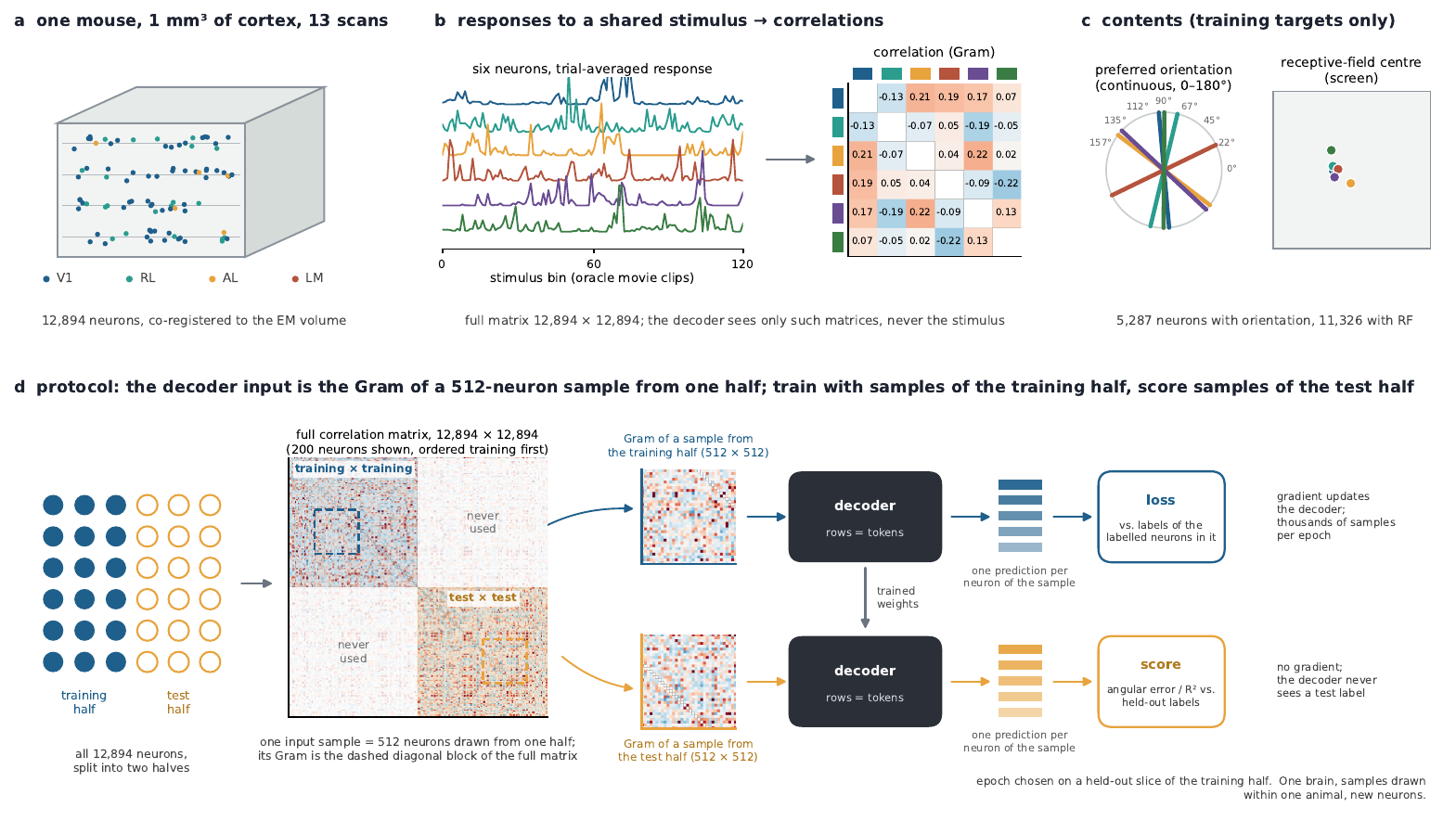}
  \caption{\textbf{How the MICrONS task is built.} (a)~One mouse, 13 of the 14 two-photon scans of a cubic millimetre of visual cortex (areas V1, RL, AL, LM), 12,894 neurons co-registered to the electron-microscopy volume. (b)~The relational substrate. Each neuron is its trial-averaged response vector to a stimulus all neurons saw (Methods); shown are six real neurons and their correlation matrix. The oracle clips were shown in all 13 scans, so the matrix is defined across scans. The decoder sees only such correlations, never the stimulus. (c)~The contents of the same six neurons: preferred orientation (in vivo, a continuous angle, 5,287 neurons with gOSI $\geq 0.25$) and receptive-field centre (digital-twin fit, 11,326 neurons). (d)~The protocol. Neurons are split into a training half and a test half; the full correlation matrix, ordered training-first (a real 200-neuron excerpt), is used only in its training$\times$training and test$\times$test blocks. A sample is 512 neurons drawn from one half; its $512\times512$ Gram (dashed) is the decoder's entire input, one row per neuron, one prediction per row. Labels enter only the loss, never the input, and the halves share no neuron (Methods).}
  \label{fig:pipeline}
\end{figure}

We answer this question with a decoder whose input is the correlation matrix of a sampled population, one row per neuron, and nothing else, and whose output is, per neuron, either a preferred orientation or a receptive-field position. It is trained on labelled populations and scored on neurons it never saw, at two scales. Within one brain we use the MICrONS release \citep{ding2025,microns2025}: 12,894 neurons from a cubic millimetre of mouse visual cortex, with in-vivo responses to a shared movie, a fitted digital-twin model \citep{wang2025}, orientation labels for 5,287 of them and receptive-field labels for 11,326 (Fig.~\ref{fig:pipeline}). The decoder trained on half the neurons reads the other half. Across brains we use the Allen Brain Observatory \citep{devries2020}: 33 mice recorded under identical stimuli. The decoder trained on some animals reads neurons of animals it never saw, which succeeds only when a sampled population mixes animals, and only with grating relations.

Averaging the correlations by orientation class shows that the relations between classes are almost circulant, which fixes the classes only up to a rotation or reflection of the orientation circle. A consistent departure from that symmetry fixes the absolute frame and makes the readout possible.

\section{Methods}

\subsection{Data}

\paragraph{MICrONS.} The node table released with \citet{ding2025} holds 12,894 neurons co-registered to the electron-microscopy volume, across V1, RL, AL and LM, from 13 of the 14 two-photon scans: the authors' working set after their quality restrictions, applied to the 13,952 matched neurons they report. Each neuron has a trial-averaged in-vivo response to the oracle natural-movie clips (120 bins), shown in every scan; a digital-twin response to a shared movie (4,999 bins, which we compress to 512 principal components; the Gram is preserved to $r > 0.999$); an in-vivo preferred orientation with a global orientation selectivity index (gOSI); and a receptive-field centre from the twin (spike-triggered-average fit, screen coordinates in $[-1,1]$). We use orientation, as a continuous angle, for the 5,287 neurons with gOSI $\geq 0.25$, and receptive-field position for the 11,326 neurons whose twin test correlation is $\geq 0.2$. The orientation labels are in-vivo measurements on both substrates; the receptive-field labels come from the twin's fit on both substrates, so every receptive-field result, in vivo included, scores relations against model-derived positions (Section~\ref{sec:discussion}). Same-scan pairs correlate more strongly than cross-scan pairs at matched receptive-field distance: with the pairs of 6,000 receptive-field-labelled neurons binned by decile of centre distance, the same-scan mean is 1.4 to 1.6 times the cross-scan mean in vivo and 1.2 to 1.4 times on the twin over the six nearest deciles. The content of the correlations is the same across scans, and all populations mix scans. Only 2\% of the receptive-field label variance lies between scans and 9\% between areas, so a decoder that read scan or area identity from the Gram could reach at most that much $\Rsq$.

\paragraph{Allen.} From Visual Coding 2P \citep{devries2020} we take the 36 VISp excitatory containers with at least 150 cells significantly responsive to drifting gratings: 33 mice, 9,281 cells, of which 5,529 carry the static-grating orientation label used here; unlabelled cells stay in the populations and receive no loss, so a held-out-animal split has 6,706 training cells and 2,575 test cells in total. Relations come from session A, either responses to natural movies shown to every mouse (4,500 bins) or the 40 blank-subtracted condition means of drifting gratings (8 directions $\times$ 5 temporal frequencies). Orientation labels come from static gratings in session B (six orientations $30^\circ$ apart, gOSI $\geq 0.25$, $p < 0.05$), so labels and relations use different stimuli. Receptive-field centres come from locally sparse noise in session C (1,504 cells with a significant fit, 11 to 112 per mouse).

\subsection{Relations}

Each neuron's response vector is centred and normalised; the relation between two neurons is the cosine of their vectors, a signal correlation. For a population of $n$ neurons the Gram is the $n\times n$ matrix of these values, standardised by the global off-diagonal mean and standard deviation, with the diagonal zeroed. We call this standardised correlation matrix the Gram throughout.

\subsection{Decoder}

Each row of a population's Gram, together with three row statistics (the mean, standard deviation and third moment of the row), is projected to a token; $L$ pre-norm transformer blocks \citep{vaswani2017,xiong2020} with $H$-head attention update the tokens, the Gram entering the attention logits as an additive bias; a linear head reads out, per token, the unit vector $(\cos 2\theta, \sin 2\theta)$ of the preferred orientation $\theta$ or an $(x,y)$ receptive-field position. The loss is the mean squared error on the labelled tokens only; unlabelled neurons stay in the population and receive no loss. The standard configuration is width 512, 8 layers and 17M parameters with populations of 512 neurons on MICrONS, and width 256, 4 layers and 2.2M parameters with populations of 256 neurons on Allen. Other sizes are in Appendix~\ref{app:scaling}. Training draws thousands of random populations per epoch.

\subsection{Protocol}

\paragraph{Split.} MICrONS neurons are halved at random. Training populations are sampled inside the training half and test populations inside the test half, so no test neuron appears in a training population (Fig.~\ref{fig:pipeline}d). Early stopping uses a slice of the training half: the epoch with the lowest error on that slice is kept, so the test half never influences model selection.

\paragraph{Regimes (Allen).} Two independent choices define a run. The split decides where test neurons come from: the training animals, each halved into training and test neurons, or held-out animals. The population decides what a sample is: neurons drawn from one animal, so that the Gram is a within-circuit matrix, or from several animals, so that most entries are between-animal correlations. We call the four cells \code{within}, \code{pooledwithin}, \code{cross} and \code{pooledcross}. Held-out regimes select the epoch on further held-out mice. Every held-out cell is run on three mouse splits and every training-animal cell on three neuron splits.

\paragraph{Metrics.} Orientation is scored by the mean absolute angular error between the decoded angle, $\tfrac12\operatorname{atan2}$ of the output, and the label, taken modulo $180^\circ$, so it lies between $0^\circ$ and $90^\circ$. A decoder whose output is independent of the label produces errors spread evenly over that range when either the labels or the outputs are uniform on the orientation circle, so chance is $45^\circ$. With the MICrONS label distribution the best constant prediction scores $41.3^\circ$ (Appendix~\ref{app:controls}). Allen labels sit on a $30^\circ$ grid, so a decoder that recovered every true preference exactly would still disagree with them by $7.5^\circ$ on average. A frame check reports the error after the best single rotation or reflection of all decoded angles. If it is lower than the raw error, the decoder recovered the structure but not the frame. Receptive fields are scored by $\Rsq$ against a baseline of 0.

\subsection{Symmetry analysis}

To expose the symmetry of the relations we average the Gram into a $K\times K$ class-Gram, the mean relation between neurons of two orientation classes, project it onto its circulant part, and ask which relabellings a held-out half's class-Gram is compatible with (all $K!$ relabellings, Frobenius distance). A relabelling that differs from the truth by a rotation or a reflection of the orientation circle is an element of the dihedral group $D_K$. The analysis reports whether the truth is singled out absolutely or only modulo $D_K$.

\section{Results}

\subsection{Per-neuron content from the correlation matrix alone}\label{sec:main}

\begin{figure}[tbp]
  \centering
  \includegraphics[width=0.85\textwidth]{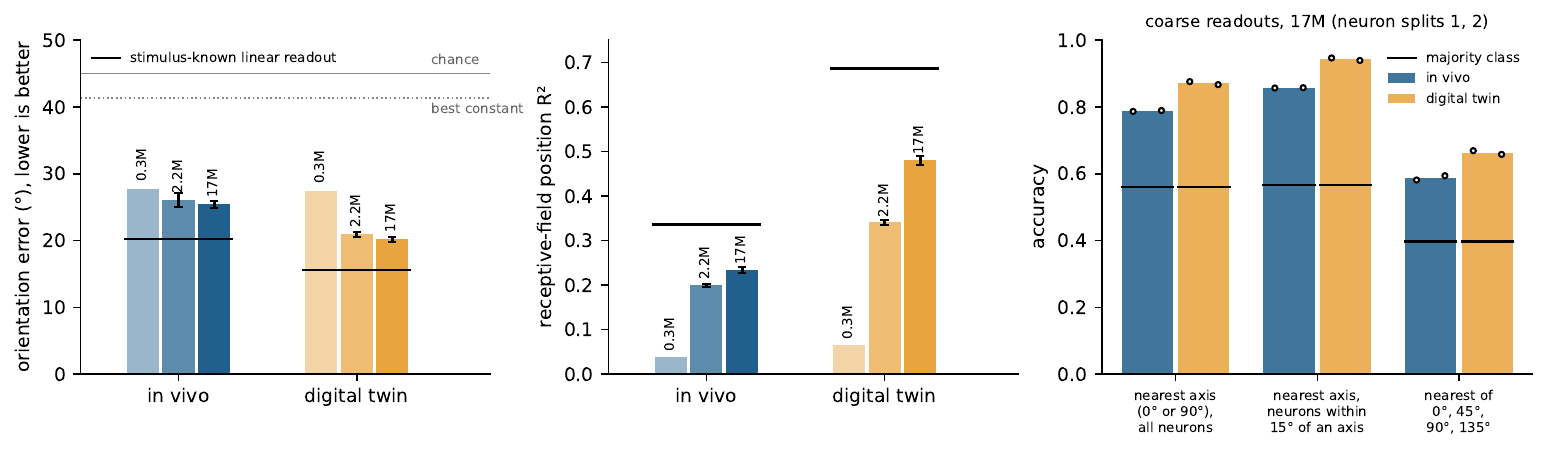}
  \caption{\textbf{MICrONS: per-neuron content decoded from the population correlation matrix alone.} Left: mean angular error of the decoded preferred orientation on held-out neurons of the same animal, for decoders of 0.3M, 2.2M and 17M parameters; error bars are the s.d.\ over 3 seeds; grey line: the $45^\circ$ error of an uninformed decoder; dotted: the best constant, $41.3^\circ$; black marks: the stimulus-known linear readout of each neuron's own responses. Middle: $\Rsq$ of the decoded receptive-field centre, same marks. Right: coarse readouts of the same decoder on neuron splits 1 and 2: whether the decoded angle falls on the correct cardinal axis, for all neurons and for the neurons whose preference lies within $15^\circ$ of an axis, and whether it falls in the correct one of four classes at $0^\circ$, $45^\circ$, $90^\circ$ and $135^\circ$; black ticks: majority class. The decoder sees only the standardised correlation matrix of 512 sampled neurons.}
  \label{fig:decoder}
\end{figure}

\begin{figure}[tbp]
  \centering
  \includegraphics[width=0.95\textwidth]{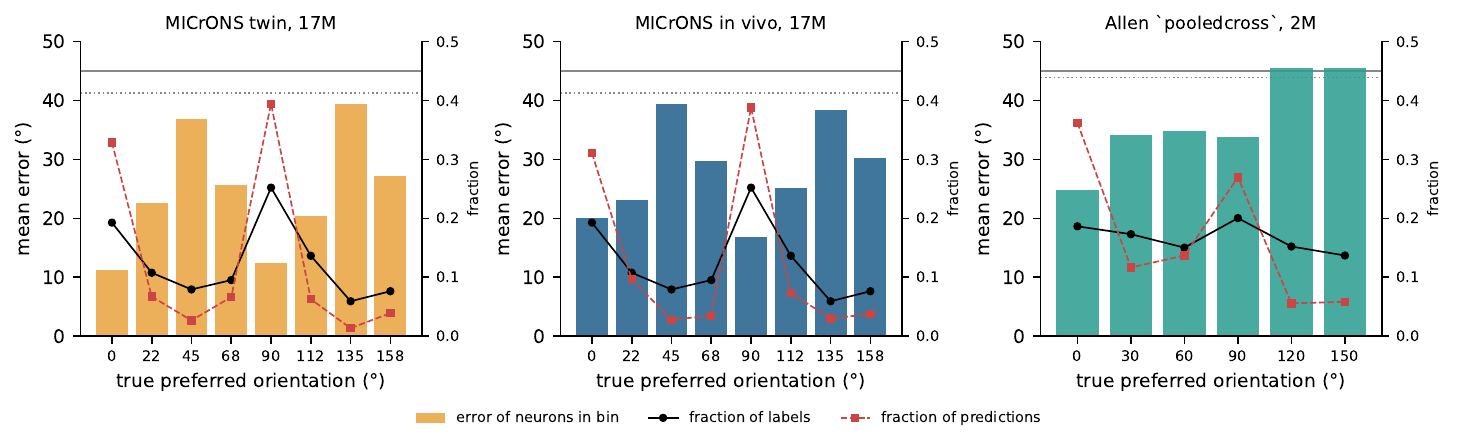}
  \caption{\textbf{The readout is cardinal.} Mean angular error of the neurons in each true-orientation bin (bars; grey line: $45^\circ$ chance; dotted: best constant), with the fraction of labels (black) and of predictions (red) per bin, for the twin and in-vivo MICrONS decoders (neuron split 2) and the Allen \code{pooledcross} decoder. Cardinal orientations are decoded well, obliques near chance, and most predictions fall on $0^\circ$ or $90^\circ$.}
  \label{fig:bins}
\end{figure}

On MICrONS the decoder reads orientation and receptive-field position from the Gram of 512 neurons of the test half (Fig.~\ref{fig:decoder}, Table~\ref{tab:microns}). The twin substrate is cleaner, since the twin removes trial noise. The predictions are in the true frame: the frame check lowers the error by less than $0.1^\circ$ in every run. About half the neurons are decoded to within $15^\circ$ (0.46 in vivo, 0.54 on the twin), but the error is not spread evenly over the orientation circle: the decoder resolves the cardinal axes and little else (Appendix~\ref{app:controls}, Fig.~\ref{fig:bins}). Asked only which cardinal axis a neuron is closer to, it is right for 87\% of all neurons on the twin and 79\% in vivo (majority answer 56\%; averaged over the eight true-orientation bins, 83\% and 74\%), and for the neurons whose preference lies within $15^\circ$ of an axis, for 94\% and 86\% (neuron splits 1 and 2). Those near-axis neurons are 56\% of the population. Sorted into four classes at $0^\circ$, $45^\circ$, $90^\circ$ and $135^\circ$, 66\% (twin) and 59\% (in vivo) land in the right one against a majority class of 40\%. Within $15^\circ$ of a cardinal axis the error is $12^\circ$ on the twin and $19^\circ$ in vivo; within $15^\circ$ of an oblique, $36^\circ$ to $39^\circ$ over both substrates, near chance (Fig.~\ref{fig:bins}). Both readouts are properties of individual neurons, not of the scan: centring predictions and labels per scan leaves the receptive-field $\Rsq$ unchanged and the orientation error within $0.2^\circ$ (Appendix~\ref{app:controls}). Test Grams from stimulus bins the decoder never trained on add about $0.5^\circ$ of error on the twin and $8.5^\circ$ in vivo, where a Gram over 60 movie bins is partly specific to those bins (Appendix~\ref{app:controls}).

\begin{table}[tbp]
  \centering\footnotesize
  \caption{MICrONS, per-neuron decoding on the test half. Relations only: the 17M decoder on the Gram of 512-neuron populations, mean $\pm$ s.d.\ over 3 seeds. Balanced: the error averaged over eight equally weighted true-orientation bins (neuron split 2). Stimulus-known linear readout: a linear map from each neuron's own movie response vector to its content, a reference that uses the stimulus alignment the relational decoder discards, not a label-noise ceiling (Appendix~\ref{app:controls}). Chance: an uninformed decoder ($45^\circ$) and the best constant orientation given the label distribution ($41.3^\circ$).}
  \label{tab:microns}
  \begin{tabular}{llcccc}
    \toprule
    content & substrate & relations only & \begin{tabular}[c]{@{}c@{}}balanced over\\orientations\end{tabular} & \begin{tabular}[c]{@{}c@{}}stimulus-known\\linear readout\end{tabular} & \begin{tabular}[c]{@{}c@{}}chance /\\best constant\end{tabular} \\
    \midrule
    orientation (angular error) & in vivo & $25.4^\circ \pm 0.5^\circ$ & $27.9^\circ$ & $20.2^\circ$ & $45^\circ$ / $41.3^\circ$ \\
    orientation (angular error) & twin & $20.2^\circ \pm 0.4^\circ$ & $24.5^\circ$ & $15.8^\circ$ & $45^\circ$ / $41.3^\circ$ \\
    receptive field ($\Rsq$) & in vivo & $0.234 \pm 0.007$ & -- & 0.34 & 0 \\
    receptive field ($\Rsq$) & twin & $0.480 \pm 0.010$ & -- & 0.68 & 0 \\
    \bottomrule
  \end{tabular}
\end{table}

\subsection{Why the absolute frame is recoverable}\label{sec:frame}

\begin{figure}[tbp]
  \centering
  \includegraphics[width=0.9\textwidth]{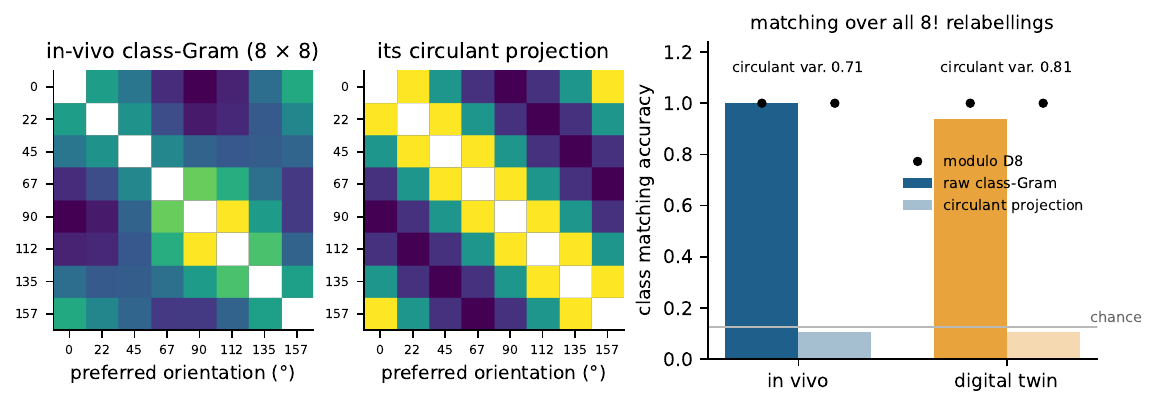}
  \caption{\textbf{What fixes the absolute orientation frame.} Left: mean correlation between orientation classes in vivo (diagonal masked). Middle: its circulant projection. Right: matching a held-out half's class-Gram to the reference over all $8!$ relabellings. With the raw matrix the true labelling wins. Accuracy is the fraction of the eight classes given the correct label. After the circulant projection it falls to 0.11 on both substrates, against the chance level of $1/8$ (grey line), while accuracy modulo $D_8$ (dots) stays at 1.0, since a circulant matrix is invariant under $D_8$: the matching then picks one of the 16 equivalent labellings, whose average per-class accuracy is $1/8$.}
  \label{fig:symmetry}
\end{figure}

The class-Gram of orientation is close to circulant (Fig.~\ref{fig:symmetry}): neighbouring orientations correlate, orthogonal ones anti-correlate, and the pattern repeats around the circle. A circulant matrix is unchanged by every rotation and reflection of the orientation circle, so that part of the relations fixes the classes only up to the 16 elements of $D_8$ and cannot say which class is $0^\circ$. Whatever fixes the frame must be a departure from circulant structure, at the class level or finer. At the class level that residual holds 29\% of the class-Gram variance in vivo and 19\% on the twin over the whole volume (30\% and 16\% in V1), and it is signal: it correlates at 0.94 between random halves of the neurons on both substrates, at 0.84 (in vivo) and 0.99 (twin) between disjoint halves of the stimulus bins, and at 0.99 with the residual computed on class-balanced subsamples, so it is neither a finite-stimulus artefact nor a consequence of the uneven number of neurons per orientation. Its form is one-peaked, not bi-cardinal: in vivo, neurons preferring orientations within about $45^\circ$ of $90^\circ$ correlate more with each other than the circulant baseline predicts and the rest less, and on both substrates the same-class excess at $90^\circ$ exceeds that at $0^\circ$ (V1: $+0.036$ against $-0.011$ in vivo, $+0.014$ against $+0.006$ on the twin, where the asymmetry is the smaller), and its shape differs between areas (residual correlation V1--RL 0.77, V1--AL 0.31).

Four observations narrow down what fixes the frame.

\emph{Priors do not explain the frame.} The decoder recovers the absolute frame (Section~\ref{sec:main}), while predictors that use only the distribution of labels, the best constant orientation or each scan's or area's mean, reach $39^\circ$ to $41^\circ$.

\emph{The structure is recoverable with no frame in the labels.} A decoder trained with each population's targets rotated at random, and the loss taken after the best alignment of predictions to targets, places neurons to $19.9^\circ$ on the twin and $25.5^\circ$ in vivo after one global alignment, the same as the standard decoder with the frame ($20.2^\circ$, $25.4^\circ$). This does not show that the circulant part alone carries the structure, since the residual is still in that decoder's input. Its raw error, $25^\circ$ and $31^\circ$ against $45^\circ$ for an arbitrary frame, shows that it fixes a consistent frame from the data even when the loss does not ask for one.

\emph{The frame is not the label density and not the particular sample of frames.} With every training and test population drawn with equal numbers of neurons from each of the eight orientation classes, the decoder reaches $22.0^\circ$ on the twin and $26.6^\circ$ in vivo (two seeds each), with no gap between raw and aligned error. The frame also survives training and test relations built from disjoint halves of the stimulus, though both halves share the movie's orientation content; that a cardinal bias of the movie is not the source is indicated by the anisotropy peaking at different orientations in areas that saw the same movie, and by Allen recovering a cardinal frame from gratings that cover orientation uniformly (Section~\ref{sec:allen}). The frame also survives removing the three row statistics from the input ($20.3^\circ$), while the statistics alone fix only a weak frame ($37.7^\circ$; Appendix~\ref{app:controls}).

\emph{The errors have the shape of the symmetry breaking.} They are $11^\circ$ to $12^\circ$ near the cardinal axes and $37^\circ$ to $39^\circ$ in the two oblique classes on the twin, and reweighting the loss so that obliques count as much as cardinals leaves them unchanged (Appendix~\ref{app:controls}). The oblique ordering is not in the relations at single-neuron resolution.

\emph{In V1 the neurons of the dominant orientation correlate most strongly with each other.} Define the anisotropy factor as the mean correlation between two neurons of the same orientation class, at the class where it peaks, divided by its mean over the other seven classes. In V1, where 28\% of neurons fall in the $22.5^\circ$ class centred on $90^\circ$, the factor peaks at $90^\circ$ on both substrates (1.28 on the twin, 1.77 in vivo; Table~\ref{tab:areas}), and the values are unchanged on class-balanced subsamples, so the signature is in the correlations, not in the counts. The coincidence therefore holds in V1, which supplies 73\% of the labelled neurons, so the volume the decoder reads inherits V1's peak; it does not hold in the two smaller areas. The quantity is a signal correlation over trial-averaged responses, and on the twin over noiseless ones, so the anisotropy is one of tuning similarity, not of trial-to-trial co-fluctuation. Cardinal over-representation and its variation between areas are established features of mouse visual cortex \citep{roth2012,fahey2019}, and the preferential connectivity of like-tuned neurons \citep{ko2011} is one route by which their responses could become more similar.

\begin{table}[htbp]
  \centering\footnotesize
  \caption{Label mode and anisotropy peak per area (MICrONS, $22.5^\circ$ classes). Label mode: the class with the most neurons and its fraction. Peak: the class with the largest mean same-class correlation and the anisotropy factor (peak over the mean of the other seven classes). LM has 10 labelled neurons and is not given its own row; the volume row includes it.}
  \label{tab:areas}
  \begin{tabular}{lcccc}
    \toprule
    area & neurons & label mode & peak, twin & peak, in vivo \\
    \midrule
    V1 & 3,858 & $90^\circ$ (0.28) & $90^\circ$ (1.28) & $90^\circ$ (1.77) \\
    RL & 1,125 & $0^\circ$ (0.24) & $112.5^\circ$ (1.22) & $112.5^\circ$ (1.68) \\
    AL & 294 & $0^\circ$ (0.32) & $0^\circ$ (1.49) & $135^\circ$ (1.69) \\
    whole volume & 5,287 & $90^\circ$ (0.25) & $90^\circ$ (1.26) & $90^\circ$ (1.81) \\
    \bottomrule
  \end{tabular}
\end{table}

\emph{A model population tests the candidate.} We could not isolate the anchor by subtraction, because removing the class-level residual from every Gram with the true classes injects the classes into the Gram (Appendix~\ref{app:controls}), so the frame is located by exclusion and by a model (Appendix~\ref{app:controls}, Table~\ref{tab:syn}). Model neurons with von Mises tuning, a movie-like stimulus that covers orientation evenly, and the same pipeline give a class-Gram circulant to within noise, and the decoder trained on it recovers the ordering, $30^\circ$ after one global alignment, but no frame, $54^\circ$ raw. Sharper cardinal tuning and a cardinal excess of neurons fix the frame in at most one seed of five, alone, together, or at any strength tested; an excess of cardinal orientations among the stimulus frames does in three of five. Adding only the anisotropy the data show, a shared stimulus-driven response component among like-tuned neurons peaked at one orientation, with uniform counts and an even stimulus, fixes the frame in every seed at every strength run (15 of 15, $2^\circ$ to $4^\circ$ raw), from a residual weaker than the twin V1's (0.05 of class-Gram variance against 0.16) to one between the twin's and the in-vivo V1 value (0.23 against 0.30); no strength tested fails, so no threshold was located; the weakest setting is below cortex on the residual fraction and above it on the anisotropy factor (1.67 against 1.28 on the twin), so the model is not calibrated to cortex in a single ordering. The model residual has the one-peaked form of the cortical one and correlates 0.39 to 0.48 with it in vivo, against 0.94 for the cortical residual with itself across neuron halves. The model is far easier to decode than cortex ($2^\circ$ against $20^\circ$ at a similar residual), so it shows that the anisotropy suffices, not that it accounts for the cortical error.

\subsection{Orientation transfers between animals through mixed populations}\label{sec:allen}

\begin{figure}[tbp]
  \centering
  \includegraphics[width=0.85\textwidth]{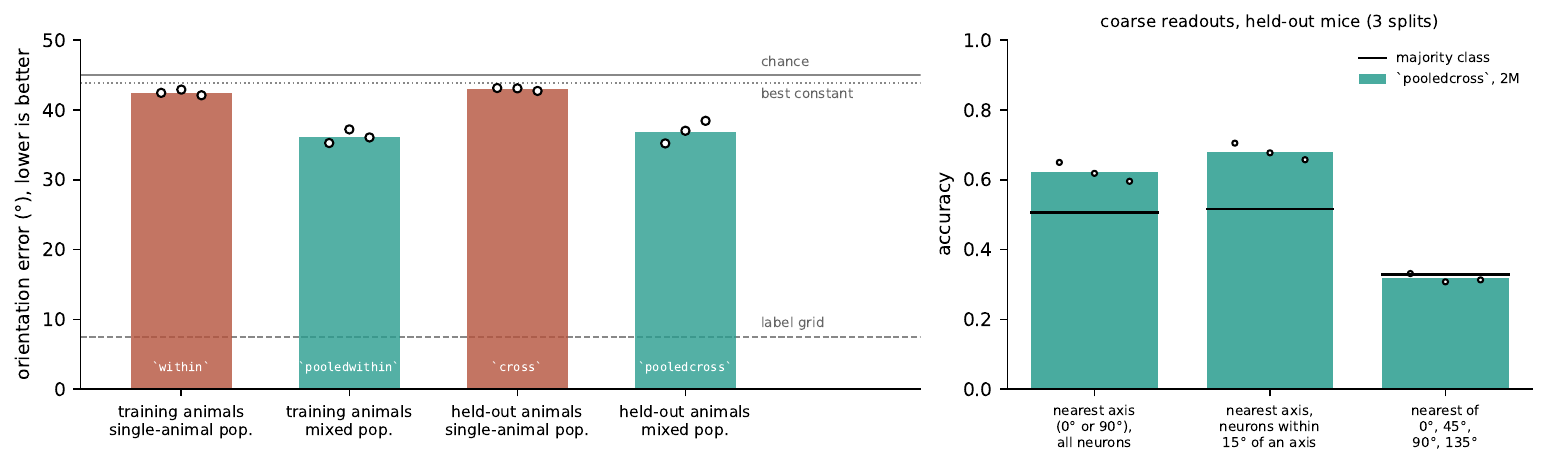}
  \caption{\textbf{Allen, 33 mice, grating relations: orientation is readable only from populations that mix animals.} Left: mean angular error; bars: mean over splits; points: the three splits; solid line: $45^\circ$ chance; dotted: $43.9^\circ$, the best constant of the split-0 held-out test set and the largest of the twelve per-test-set values; each run's own test set has a best constant of $42.5^\circ$ to $43.9^\circ$, so the single-animal bars are not below their own baselines; dashed: the $7.5^\circ$ label grid. Single-animal populations are $1.9^\circ$ to $2.9^\circ$ below chance whether the animal was seen in training or not; mixed populations are $6.6^\circ$ to $9.8^\circ$ below it in both rows. Right: coarse readouts of the mixed-population decoder on the held-out mice (three splits): correct cardinal axis for all cells and for the cells labelled $0^\circ$ or $90^\circ$; the third bar, nearest of four classes $45^\circ$ apart, maps each $30^\circ$ label to one class ($30^\circ$ and $60^\circ$ to $45^\circ$, $120^\circ$ and $150^\circ$ to $135^\circ$) and is shown for comparison with MICrONS; black ticks: majority class.}
  \label{fig:allen2x2}
\end{figure}

With natural-movie relations, orientation is nearly absent from Allen correlations: same-orientation pairs correlate 0.005 more than orthogonal pairs, against 0.09 in MICrONS, and the class-level tests are at or just above chance (Fig.~\ref{fig:a4}). The Allen results below therefore use grating relations, a designed set of 40 conditions over an orientation grid, and do not replicate on naturalistic relations. With gratings the signal is present (0.05 within a mouse, 0.044 across mice) and the four regimes separate (Fig.~\ref{fig:allen2x2}, Table~\ref{tab:allen}). Single-animal populations are close to uninformative: all six runs lie $1.9^\circ$ to $2.9^\circ$ below the $45^\circ$ uninformed level, but five of the six are only $0.3^\circ$ to $1.0^\circ$ below the best constant of their own test set ($42.5^\circ$ to $43.9^\circ$ per split) and the sixth is $0.4^\circ$ above it. A margin of that size is inside the sampling noise of a test set of about 1,500 cells (bootstrap s.e.\ $0.9^\circ$ over cells, $1.4^\circ$ clustered on the nine test mice), so the six runs show a consistent direction, not a resolved effect, whether or not the animal was seen in training, and on every split the mixed populations of the same regime are $4.3^\circ$ to $7.9^\circ$ better (mixed populations of 128 neurons, matching the single-animal size, score $36.0^\circ$; Appendix~\ref{app:controls}). Mixed populations are $6.6^\circ$ to $9.8^\circ$ below chance in both rows, at the same level, so no loss from transfer to a new brain is detectable at three splits. As on MICrONS the readout is cardinal: over the three splits the decoder names the right cardinal axis for 62\% of all cells and 68\% of the cells labelled $0^\circ$ or $90^\circ$ (majority 51\% and 52\%; 61\% averaged over the six label bins), cells labelled $0^\circ$ are decoded to $25^\circ$ and those labelled $90^\circ$ to $34^\circ$, cells labelled $120^\circ$ or $150^\circ$ are at chance, and the four-class readout is at the majority class (Fig.~\ref{fig:bins}). What the mixed populations add is neurons from other animals, both as entries of the Gram and as diversity in the pool a population is drawn from; the controls below separate the two only partly. These exist only because all mice saw the same 40 conditions, and they measure how similar two tuning profiles are. The decoder still never sees which condition is which, so the absolute frame must again come from the population's relational structure. Orientation is thus readable from a correlation geometry that spans animals, and only weakly from relations inside one circuit. Two controls say why. The limit is not the neurons: with each neuron's own 40-condition response vector as its token, the same decoder on the same single-animal populations reaches $33^\circ$ to $35^\circ$, as with mixed populations (Table~\ref{tab:allen}, last row). The limit is not the total number of training cells either: on held-out animals the single-animal and the mixed runs are trained on exactly the same cells (6,706, 6,927 and 6,853 by split), and on the training animals on 3,451 to 3,734 against 3,480; only the sampling of each population is confined to one animal. An animal supplies 171 to 633 cells, 84 to 395 of them labelled, and a mouse whose training half holds fewer than the 128 neurons a population needs is skipped by the sampler, so the populations of one animal overlap in a large fraction of their neurons. Shrinking the pool instead, with mixed populations of 128 drawn from 280 or 560 training cells, costs $4.5^\circ$ ($40.5^\circ$ and $40.7^\circ$ against $36.0^\circ$ from the full pool of 6,706; single runs, Appendix~\ref{app:controls}), less than the $7.9^\circ$ that confining populations to one animal costs on the same split (held-out split 0), and those mixed populations still span animals, so no run here isolates the between-animal entries themselves. MICrONS separates pool size from confinement: a decoder trained on 140 neurons of one scan and tested on the other scans scores $38^\circ$ to $44^\circ$ on the twin and $37^\circ$ to $39^\circ$ in vivo (three scans), while with the full training pool of one animal, single-scan populations of 128 are still decoded only to $32.0^\circ$ on the twin and $31.1^\circ$ in vivo, against $22.3^\circ$ and $25.8^\circ$ for mixed-scan populations of the same size (three seeds each; Appendix~\ref{app:controls}). Confining populations to one scan therefore costs $5^\circ$ to $10^\circ$ even with the full training pool.

\begin{table}[tbp]
  \centering\footnotesize
  \caption{Allen, mean angular error of the decoded orientation in the four regimes (2.2M decoder, 256-neuron populations; 128 for single-animal populations), three splits per cell (of neurons within each animal for the top row, of animals for the second row). Last row: the same decoder on held-out animals with each neuron's 40-condition response vector as its token instead of its Gram row, the stimulus-known reference. Chance $45^\circ$; best constant of each run's own test set $42.5^\circ$ to $43.9^\circ$ ($43.7^\circ$ over all 5,529 labelled cells); label grid $7.5^\circ$. Means over the three splits in parentheses.}
  \label{tab:allen}
  \begin{tabular}{lcc}
    \toprule
    test neurons from & single-animal populations & mixed populations \\
    \midrule
    the training animals & $42.5^\circ$ / $42.9^\circ$ / $42.1^\circ$ ($42.5^\circ$) & $35.3^\circ$ / $37.2^\circ$ / $36.1^\circ$ ($36.2^\circ$) \\
    held-out animals & $43.1^\circ$ / $43.1^\circ$ / $42.7^\circ$ ($43.0^\circ$) & $35.2^\circ$ / $37.0^\circ$ / $38.4^\circ$ ($36.9^\circ$) \\
    \midrule
    held-out animals, stimulus-known reference & $32.8^\circ$ / $33.6^\circ$ / $34.9^\circ$ ($33.8^\circ$) & $32.4^\circ$ / $33.7^\circ$ / $34.3^\circ$ ($33.5^\circ$) \\
    \bottomrule
  \end{tabular}
\end{table}

\subsection{Receptive fields: what crosses animals is the animal's screen position}\label{sec:rf}

\begin{figure}[tbp]
  \centering
  \includegraphics[width=0.85\textwidth]{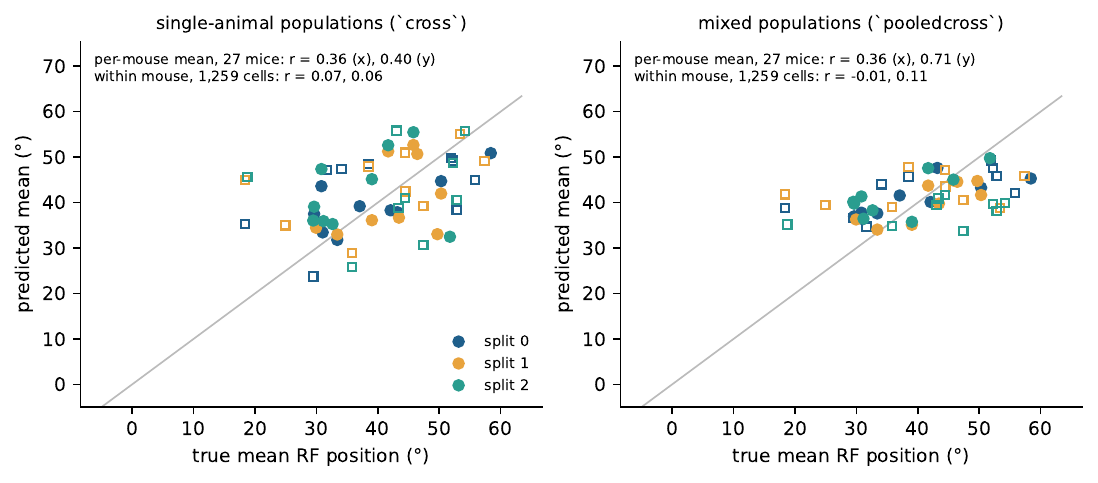}
  \caption{\textbf{Allen receptive fields: what crosses animals is each animal's screen position.} Predicted versus true mean receptive-field position of each held-out mouse (27 points over 3 splits; $x$ squares, $y$ dots), from decoders trained on single-animal (left) or mixed (right) populations. The per-mouse mean is recovered; the position of a neuron relative to its mouse-mates is not.}
  \label{fig:rf}
\end{figure}

Correlation falls with receptive-field distance within a mouse (0.127 at $<5^\circ$ to 0.068 beyond $40^\circ$) and across mice (0.049 to 0.011). Decoding absolute screen coordinates on held-out mice gives $\Rsq$ that varies with the split, from $0.29$ to $-0.45$ for single-animal populations and from $0.22$ to $-0.25$ for mixed ones. Decomposing the saved predictions explains the variation (Fig.~\ref{fig:rf}, Appendix~\ref{app:rfsplits}). The correlation between predicted and true mean position of each held-out mouse, pooled over 27 mouse-level predictions, is 0.36 to 0.71 (permutation $p \leq 0.03$). The correlation between predicted and true position of a neuron relative to its mouse-mates is $-0.01$ to 0.11 over 1,259 cells. The decoder recovers where an animal's imaged patch looks, not which neuron within it looks where. $\Rsq$ on absolute coordinates then depends mostly on whether the between-mouse spread of a particular test set is reproduced at the right scale.

The within-mouse layout is present in the correlations, since same-mouse correlation falls with distance. It is not learned; each mouse contributes only 11 to 112 labelled cells. With 11,326 labels in one animal, MICrONS lets the same architecture reach neuron-level receptive fields at $\Rsq = 0.48$ with the 17M decoder, and 0.34 at the Allen decoder's size of 2.2M.

\clearpage
\section{Discussion}\label{sec:discussion}

The correlation matrix of a few hundred cortical neurons, with the stimulus unknown and no neuron labelled, determines each neuron's preferred orientation to a mean error of $25^\circ$ in vivo and $20^\circ$ on the digital twin, against $45^\circ$ for an uninformed decoder and $41^\circ$ for the best constant guess. The readout is cardinal: near-cardinal neurons to $12^\circ$ to $19^\circ$, obliques near chance. It determines each neuron's receptive-field position to $\Rsq = 0.23$ in vivo and 0.48 on the digital twin. The symmetry analysis constrains the mechanism: the relations fix the class structure up to the symmetries of the content, and a departure from that symmetry fixes the frame. That departure is a property of the correlations themselves: it is not the uneven number of neurons per orientation and not the particular stimulus sample. In V1 the neurons preferring the over-represented orientation have the most similar responses to one another, whether or not the sampled population is imbalanced. A direct ablation of that feature in the data is not available, but a model population with that anisotropy and nothing else recovers the absolute frame at every strength tested, while models with sharper or more numerous cardinal neurons do not.

Across animals the two contents behave differently. Orientation is carried by the relations between tuning profiles, which exist between any two neurons that saw the same conditions, in the same brain or not. From the correlations inside a single scan it is readable only weakly, at $31^\circ$ to $32^\circ$ on MICrONS, and inside a single Allen animal not measurably at all: each animal supplies 171 to 633 cells, so its populations of 128 neurons overlap in a large fraction of their neurons and differ too little from one another. Receptive-field position is carried at the level of an animal's patch, and the neuron-level layout needs more labels per animal than these recordings provide.

Limits. MICrONS is one animal, and its training and test halves share recording sessions. The Allen held-out-animal cells are the cleaner generalisation test, but their variance across mouse splits is large, and they exist only with grating relations, a designed 40-condition set on which two tuning profiles can be compared, not with the naturalistic relations of the MICrONS result. The readout is cardinal: the obliques are not resolved on either dataset, so the mean error understates the precision at the cardinals and overstates it elsewhere. No experiment removes the residual from the decoder's input, so the assignment of the ordering to the circulant part and the frame to the residual is inferred from the symmetry argument and the model, not measured. The mechanism is established by exclusion and by a model, not by removing the candidate anisotropy from cortical data, since the direct ablation we tried injected the labels it was meant to remove (Appendix~\ref{app:controls}). Orientation labels are noisy in both datasets (in vivo and twin labels agree on 70\% of MICrONS neurons at 8 classes; static and drifting gratings on 45 to 67\% of Allen cells), and the Allen labels are quantised to $30^\circ$, which caps every labelled and label-free number alike. The twin substrate is a model fitted to the same recordings, so the in-vivo results carry the claim about cortex and the twin results show what cleaner correlations would give; in AL the two substrates place the anisotropy peak two classes apart ($0^\circ$ on the twin, $135^\circ$ in vivo), though the in-vivo peak is within noise of three other classes at 294 neurons. The receptive-field labels come from the twin for both substrates, so the twin receptive-field result compares a model's correlations with the same model's receptive fields, and the in-vivo result scores cortical correlations against model-estimated positions. The orientation labels are in-vivo measurements throughout.

The symmetry result is an instance of a general expectation: a representation that is equivariant to a group has a group-fixed part, identical across individuals up to relabelling, and an individual remainder \citep{oizumi2026}. The matching-over-relabellings test is the discrete form of unsupervised alignment of similarity structures across individuals by optimal transport \citep{kawakita2025}. This paper adds the measurement in cortex, at the level of single neurons, and the finding that the absolute readout is possible and rests on a remainder that belongs to the correlations, not to the sample of neurons or stimuli.

\paragraph{Code and data.} Code, run configurations and per-neuron predictions are in the directory \texttt{microns\_ambiguity} (branch \texttt{microns-ambiguity}) of \url{https://github.com/entropicbloom/intentionality}. All data are public releases.

\paragraph{Use of language models.} Language-model assistance (Claude, Anthropic) was used for drafting and editing the text, for analysis code, and for internal review passes. The study design, analyses and conclusions are the author's, who verified results, statements and references.

\bibliographystyle{unsrtnat}
\bibliography{refs}

\begin{thebibliography}{18}
\providecommand{\natexlab}[1]{#1}
\providecommand{\url}[1]{\texttt{#1}}
\expandafter\ifx\csname urlstyle\endcsname\relax
  \providecommand{\doi}[1]{doi: #1}\else
  \providecommand{\doi}{doi: \begingroup \urlstyle{rm}\Url}\fi

\bibitem[Shepard and Chipman(1970)]{shepard1970}
Roger~N. Shepard and Susan Chipman.
\newblock Second-order isomorphism of internal representations: shapes of
  states.
\newblock \emph{Cognitive Psychology}, 1:\penalty0 1--17, 1970.

\bibitem[Kriegeskorte et~al.(2008)Kriegeskorte, Mur, and
  Bandettini]{kriegeskorte2008}
Nikolaus Kriegeskorte, Marieke Mur, and Peter~A. Bandettini.
\newblock Representational similarity analysis: connecting the branches of
  systems neuroscience.
\newblock \emph{Frontiers in Systems Neuroscience}, 2:\penalty0 4, 2008.

\bibitem[Lyre(2022)]{lyre2022}
Holger Lyre.
\newblock Neurophenomenal structuralism. {A} philosophical agenda for a
  structuralist neuroscience of consciousness.
\newblock \emph{Neuroscience of Consciousness}, 2022\penalty0 (1):\penalty0
  niac012, 2022.

\bibitem[L{\"a}ssig(2025)]{framework2025}
Francesco L{\"a}ssig.
\newblock Unambiguous representations in neural networks: an
  information-theoretic approach to intentionality.
\newblock arXiv:2512.11000 [q-bio.NC], 2025.
\newblock URL \url{https://arxiv.org/abs/2512.11000}.

\bibitem[Tsodyks et~al.(1999)Tsodyks, Kenet, Grinvald, and Arieli]{tsodyks1999}
Misha Tsodyks, Tal Kenet, Amiram Grinvald, and Amos Arieli.
\newblock Linking spontaneous activity of single cortical neurons and the
  underlying functional architecture.
\newblock \emph{Science}, 286:\penalty0 1943--1946, 1999.

\bibitem[Rubin et~al.(2019)Rubin, Sheintuch, Brande-Eilat, Pinchasof, Rechavi,
  Geva, and Ziv]{rubin2019}
Alon Rubin, Liron Sheintuch, Noa Brande-Eilat, Or~Pinchasof, Yoav Rechavi,
  Nitzan Geva, and Yaniv Ziv.
\newblock Revealing neural correlates of behavior without behavioral
  measurements.
\newblock \emph{Nature Communications}, 10:\penalty0 4745, 2019.

\bibitem[Arora et~al.(2025)Arora, Lachi, Knight, Azabou, Richards, Hurwitz,
  Siegle, and Dyer]{arora2025}
Vinam Arora, Divyansha Lachi, Ian~J. Knight, Mehdi Azabou, Blake Richards,
  Cole~L. Hurwitz, Joshua~H. Siegle, and Eva~L. Dyer.
\newblock Know thyself by knowing others: learning neuron identity from
  population context.
\newblock In \emph{Advances in Neural Information Processing Systems},
  volume~38, 2025.

\bibitem[Ding et~al.(2025)Ding, Fahey, Papadopoulos, et~al.]{ding2025}
Zhuokun Ding, Paul~G. Fahey, Stelios Papadopoulos, et~al.
\newblock Functional connectomics reveals general wiring rule in mouse visual
  cortex.
\newblock \emph{Nature}, 640\penalty0 (8058):\penalty0 459--469, 2025.

\bibitem[{The MICrONS Consortium}(2025)]{microns2025}
{The MICrONS Consortium}.
\newblock Functional connectomics spanning multiple areas of mouse visual
  cortex.
\newblock \emph{Nature}, 640\penalty0 (8058):\penalty0 435--447, 2025.

\bibitem[Wang et~al.(2025)Wang, Fahey, Ding, et~al.]{wang2025}
Eric~Y. Wang, Paul~G. Fahey, Zhuokun Ding, et~al.
\newblock Foundation model of neural activity predicts response to new stimulus
  types.
\newblock \emph{Nature}, 640\penalty0 (8058):\penalty0 470--477, 2025.

\bibitem[{de Vries} et~al.(2020){de Vries}, Lecoq, Buice, et~al.]{devries2020}
Saskia E.~J. {de Vries}, Jerome~A. Lecoq, Michael~A. Buice, et~al.
\newblock A large-scale standardized physiological survey reveals functional
  organization of the mouse visual cortex.
\newblock \emph{Nature Neuroscience}, 23:\penalty0 138--151, 2020.

\bibitem[Vaswani et~al.(2017)Vaswani, Shazeer, Parmar, Uszkoreit, Jones, Gomez,
  Kaiser, and Polosukhin]{vaswani2017}
Ashish Vaswani, Noam Shazeer, Niki Parmar, Jakob Uszkoreit, Llion Jones,
  Aidan~N. Gomez, {\L}ukasz Kaiser, and Illia Polosukhin.
\newblock Attention is all you need.
\newblock In \emph{Advances in Neural Information Processing Systems},
  volume~30, 2017.

\bibitem[Xiong et~al.(2020)Xiong, Yang, He, Zheng, Zheng, Xing, Zhang, Lan,
  Wang, and Liu]{xiong2020}
Ruibin Xiong, Yunchang Yang, Di~He, Kai Zheng, Shuxin Zheng, Chen Xing,
  Huishuai Zhang, Yanyan Lan, Liwei Wang, and Tie-Yan Liu.
\newblock On layer normalization in the transformer architecture.
\newblock In \emph{International Conference on Machine Learning}, volume 119 of
  \emph{PMLR}, pages 10524--10533, 2020.

\bibitem[Roth et~al.(2012)Roth, Helmchen, and Kampa]{roth2012}
Morgane~M. Roth, Fritjof Helmchen, and Bj{\"o}rn~M. Kampa.
\newblock Distinct functional properties of primary and posteromedial visual
  area of mouse neocortex.
\newblock \emph{Journal of Neuroscience}, 32\penalty0 (28):\penalty0
  9716--9726, 2012.

\bibitem[Fahey et~al.(2019)Fahey, Muhammad, Smith, Froudarakis, Cobos, Fu,
  Walker, Yatsenko, Sinz, Reimer, and Tolias]{fahey2019}
Paul~G. Fahey, Taliah Muhammad, Cameron Smith, Emmanouil Froudarakis, Erick
  Cobos, Jiakun Fu, Edgar~Y. Walker, Dimitri Yatsenko, Fabian~H. Sinz, Jacob
  Reimer, and Andreas~S. Tolias.
\newblock A global map of orientation tuning in mouse visual cortex, 2019.
\newblock bioRxiv 745323.

\bibitem[Ko et~al.(2011)Ko, Hofer, Pichler, Buchanan, Sj{\"o}str{\"o}m, and
  Mrsic-Flogel]{ko2011}
Ho~Ko, Sonja~B. Hofer, Bruno Pichler, Katherine~A. Buchanan, P.~Jesper
  Sj{\"o}str{\"o}m, and Thomas~D. Mrsic-Flogel.
\newblock Functional specificity of local synaptic connections in neocortical
  networks.
\newblock \emph{Nature}, 473:\penalty0 87--91, 2011.

\bibitem[Oizumi et~al.(2026)Oizumi, Lim, and Kanai]{oizumi2026}
Masafumi Oizumi, Chanseok Lim, and Ryota Kanai.
\newblock Principal bundle geometry of qualia: understanding the quality of
  consciousness from symmetry.
\newblock \emph{PNAS Nexus}, 5\penalty0 (9):\penalty0 pgag261, 2026.

\bibitem[Kawakita et~al.(2025)Kawakita, Zeleznikow-Johnston, Takeda, Tsuchiya,
  and Oizumi]{kawakita2025}
Genji Kawakita, Ariel Zeleznikow-Johnston, Ken Takeda, Naotsugu Tsuchiya, and
  Masafumi Oizumi.
\newblock Is my ``red'' your ``red''?: {E}valuating structural correspondences
  between color similarity judgments using unsupervised alignment.
\newblock \emph{iScience}, 28\penalty0 (3):\penalty0 112029, 2025.

\end{thebibliography}

\clearpage
\appendix
\renewcommand{\thefigure}{A\arabic{figure}}
\setcounter{figure}{0}
\section{Scaling and population size}\label{app:scaling}
\begin{figure}[p]
  \centering
  \includegraphics[width=\textwidth]{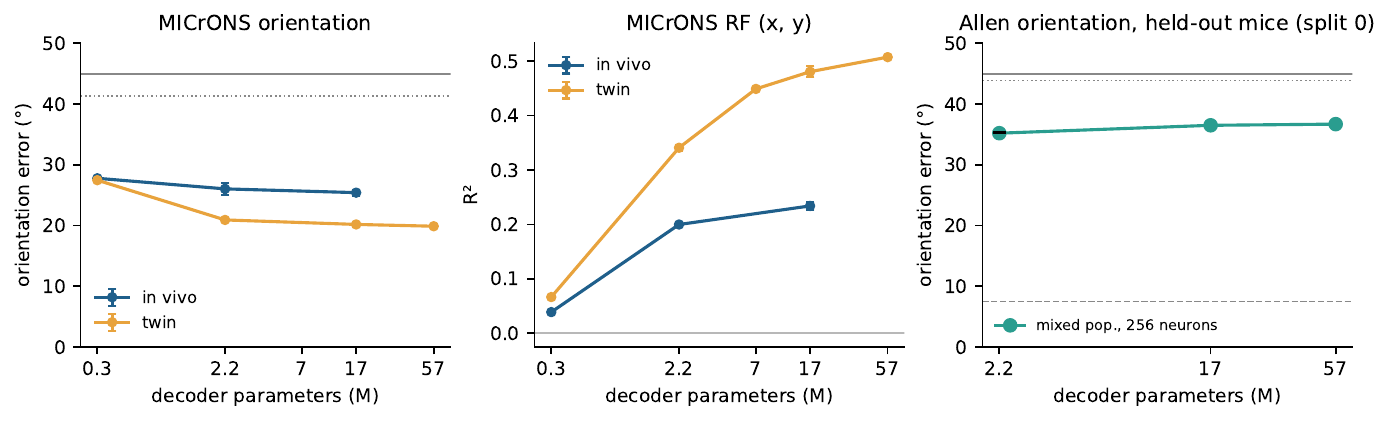}
  \caption{\textbf{Scaling with decoder size.} Test metric versus number of decoder parameters at fixed population size (512 neurons MICrONS, 256 Allen). Error bars: s.d.\ over 3 seeds where run. On MICrONS receptive fields the gain continues to 57M; the orientation error is flat from 2.2M upward on both substrates and on Allen.}
  \label{fig:a1}
\end{figure}

\begin{figure}[p]
  \centering
  \includegraphics[width=0.85\textwidth]{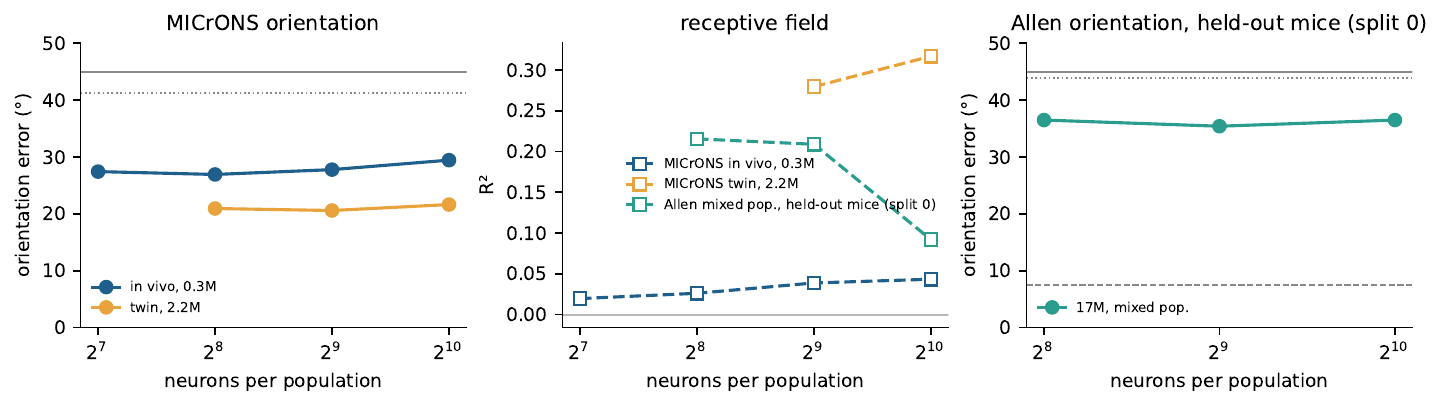}
  \caption{\textbf{Population size.} Test metric versus the number of neurons in each sampled population. Beyond 256 to 512 neurons the orientation error moves by at most $2^\circ$, up or down; twin receptive fields keep gaining slowly, Allen receptive fields lose. The twin 2.2M series was run at 512 and 1,024 neurons and the Allen series from 256 upward; only the in-vivo 0.3M series includes 128.}
  \label{fig:a2}
\end{figure}

\begin{figure}[p]
  \centering
  \includegraphics[width=\textwidth]{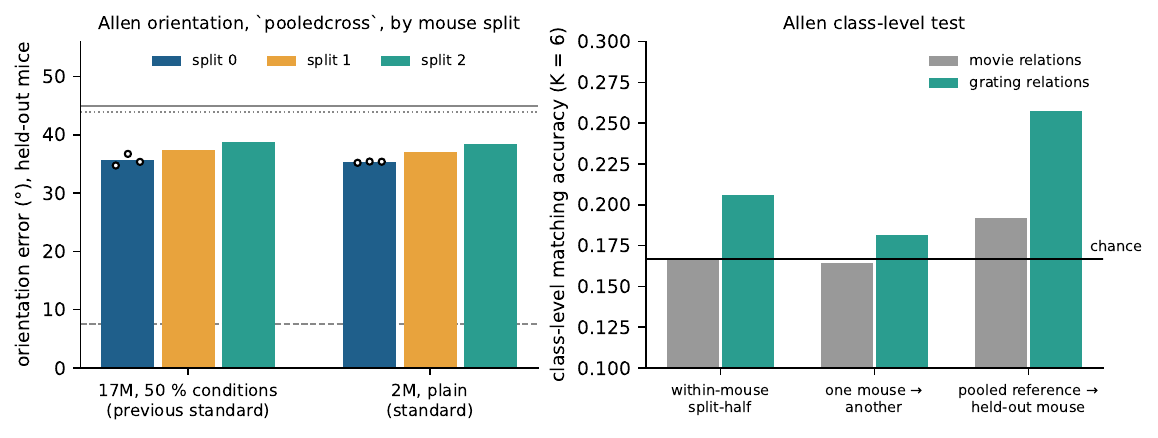}
  \caption{\textbf{Allen split variance and the class-level test.} Left: \code{pooledcross} orientation error on the same three held-out-mouse splits for the previous standard configuration (17M, with a training-time Gram augmentation no longer used) and the final one (2.2M); split 0 carries three seeds. Right: the class-level test with relations from natural movies versus drifting gratings; only the grating relations move every test off chance.}
  \label{fig:a4}
\end{figure}

More capacity helps where labels are plentiful and not where they are scarce (Fig.~\ref{fig:a1}). Twin receptive fields go from 0.34 to 0.48 to 0.51 $\Rsq$ between 2.2M, 17M and 57M parameters. Orientation error is flat from 2.2M upward on both MICrONS substrates ($20.9^\circ$, $20.2^\circ$, $19.9^\circ$ on the twin at 2.2M, 17M and 57M) and on Allen ($35.2^\circ$, $36.5^\circ$, $36.7^\circ$); only the 0.3M model is worse ($27.5^\circ$ on the twin and $27.8^\circ$ in vivo). Every model above 2M memorises its training neurons within 10 to 20 epochs and is early-stopped on the held-out selection slice. Population size beyond 256 to 512 neurons changes orientation error by at most $2^\circ$ in either direction, while twin receptive fields keep gaining slowly (Fig.~\ref{fig:a2}). Split-to-split variance on held-out mice exceeds seed variance (Fig.~\ref{fig:a4}).

\section{Controls}
\label{app:controls}
Controls on the MICrONS standard configuration (17M, 512-neuron populations, digital twin unless stated; Table~\ref{tab:controls}).

\paragraph{What the decoder reads.} With the three row statistics as the only input per neuron and no relational term, the error is $37.7^\circ$, raw and after alignment alike, so the scalars fix a weak frame on their own; with the Gram row alone and no statistics it is $20.3^\circ$ on the twin and $26.5^\circ$ in vivo, raw and aligned, so the relations carry the frame and the scalars are dispensable. (They are not a selectivity confound: their correlations with distance to the nearest cardinal axis are below 0.12 in magnitude and with gOSI below 0.16.) With the statistics plus the Gram entering only as an attention bias, which makes the decoder equivariant to the order of the neurons, the error is $19.9^\circ$ over three seeds, the same as the standard decoder ($20.2^\circ$), whose input layer also projects the raw Gram row in sample order.

\paragraph{Reference decoders that see the activity.} To measure what the relations discard, the same transformer was trained with each neuron's token built from its response vector instead of its Gram row, so that it sees the stimulus-aligned activity the Gram is computed from (Table~\ref{tab:ref}). Adding the Gram to it as an attention bias changes nothing. A linear readout of each neuron's own response vector, with no population context, does as well as the activity transformer, so the activity decoder is a per-neuron tuning readout whose only advantage over the relational decoder is the alignment of the responses to the stimulus. On orientation the relations retain about 80\% of the gain over the best constant that the aligned activity gives, and on receptive fields about 70\%; the twin receptive-field gap is inflated by the twin's labels and responses coming from the same model. With test responses from movie frames the decoder never trained on, the activity decoder has no valid input, since its coordinates now index different frames, and it scores at chance or worse, while the relational decoder scores $33.8^\circ$ in vivo and $20.7^\circ$ on the twin. This shows which inputs each readout can be applied to, not which readout is better. The activity decoder transfers across animals under a shared stimulus protocol (Section~\ref{sec:allen}) but not across stimuli. The relations do both, because a correlation does not depend on which frames it was computed from, only on both neurons having seen the same ones. A decoder given the response vectors with the stimulus bins shuffled, the same shuffle for every neuron of a population, is blind to the stimulus but free to compute any relation among the neurons. It is no better than the Gram decoder on orientation and far below it on receptive fields (Table~\ref{tab:ref}), but it has to discover the relations under a fresh permutation for every population, so the comparison does not show whether richer statistics than the pairwise Gram exist.

\begin{table}[htbp]
  \centering\footnotesize
  \caption{Reference decoders that see the activity (each neuron's stimulus-aligned response vector), 17M unless linear, 512-neuron populations. Orientation as mean angular error (chance $45^\circ$), receptive field as $\Rsq$; mean $\pm$ s.d.\ over three seeds where given, otherwise seed 0.}
  \label{tab:ref}
  \begin{tabular}{lcccc}
    \toprule
    decoder input & ori.\ in vivo & ori.\ twin & RF in vivo & RF twin \\
    \midrule
    Gram row (this paper) & $25.4^\circ \pm 0.5^\circ$ & $20.2^\circ \pm 0.4^\circ$ & $0.23 \pm 0.01$ & $0.48 \pm 0.01$ \\
    activity, transformer & $21.5^\circ \pm 0.4^\circ$ & $15.0^\circ \pm 0.1^\circ$ & 0.37 & 0.80 \\
    activity + Gram as attention bias & $22.2^\circ$ & $15.0^\circ$ & -- & -- \\
    activity, linear, no population context & $20.2^\circ \pm 0.2^\circ$ & $15.8^\circ \pm 0.1^\circ$ & $0.34 \pm 0.01$ & $0.68 \pm 0.01$ \\
    activity, transformer, disjoint stimulus bins & $49.5^\circ$ & $41.0^\circ$ & -- & -- \\
    activity, bins shuffled per population, transformer & $26.3^\circ \pm 0.6^\circ$ & $27.8^\circ \pm 6.5^\circ$ & 0.04 & 0.08 \\
    activity, bins shuffled, linear, no population context & $40.3^\circ$ & $40.2^\circ$ & -- & -- \\
    \bottomrule
  \end{tabular}
\end{table}

\paragraph{Cross-stimulus test.} Training and selection Grams built from one random half of the stimulus bins and test Grams from the other half give $20.7^\circ$ on the twin (2,500 bins per half, raw responses rather than principal components; three seeds, 20.3 to 21.2) and $33.8^\circ$ in vivo (60 bins per half; 33.4 to 34.2), against $20.2^\circ$ and $25.4^\circ$ with shared bins. A control with the same 60-bin half on both sides gives $25.3^\circ$, so the in-vivo increase comes from the stimulus sample, not the smaller Gram: a correlation matrix over 60 movie bins partly reflects which frames it was built from, and one over 2,500 bins does not.

\paragraph{Scan and area.} The MICrONS neurons come from 13 scans in four areas, same-scan pairs correlate more strongly than cross-scan pairs, and the scans look at different parts of the screen, so a decoder that read scan identity from the Gram could predict each scan's mean receptive-field position and score a positive $\Rsq$ without knowing any neuron's position. The Allen decoder does exactly that (Section~\ref{sec:rf}), so we checked MICrONS the same way, centring predictions and labels per scan and per area before scoring. The receptive-field result holds: on a fourth seed of the Table~\ref{tab:microns} configuration, run with saved predictions, $\Rsq$ is 0.499 absolute and 0.498 within scan on the twin, 0.249 and 0.253 in vivo, and 0.47 (twin) and 0.22 (in vivo) within area. Only 2\% of the label variance lies between scans and 9\% between areas, so a scan-identity readout could reach at most 0.02. For orientation, a predictor that assigns every neuron its scan's or area's mean orientation scores $39^\circ$, and removing a separate rotation per scan or area changes the decoder's error by less than $0.2^\circ$. Decoders trained on V1 alone transfer to the other areas unevenly across seeds (Table~\ref{tab:controls}): to AL the aligned error is $32^\circ$ in all three seeds but only one seed lands on a rotated frame, and to RL one seed transfers and two do not. The RL row uses 512-neuron populations and the V1 and AL rows 128. With class-balanced training populations, V1 itself is decoded $9^\circ$ worse than with the standard training set, against $1.8^\circ$ for class balancing of the whole volume, because V1's cardinal-heavy test set rewards the prior that balancing removes.

\paragraph{Class balance and priors.} Preferred orientations are uneven in MICrONS (19\% of labelled neurons within $11^\circ$ of $0^\circ$, 25\% of $90^\circ$, 6 to 10\% in each oblique bin), so a decoder could gain from the distribution alone. It gains little: the best constant prediction has an error of $41.3^\circ$, a predictor that assigns every neuron the circular mean of its scan $39.0^\circ$ and of its area $38.7^\circ$, against $20.3^\circ$ for the decoder on the twin and $24.6^\circ$ in vivo (neuron split 2). Training and testing on populations drawn with equal numbers of neurons from each of the eight classes (64 per class) gives $22.0^\circ$ on the twin and $26.6^\circ$ in vivo (two seeds: 22.4, 21.5; 26.5, 26.7), raw and after global alignment alike, so the absolute frame does not depend on the class imbalance. The per-area errors are within $3^\circ$ of each other: the frame is read from a volume-wide sample, so an area whose label mode differs from the volume's peak is not mis-framed. On Allen the labels are nearly flat (14 to 20\% per orientation), the best constant of that test set is $43.9^\circ$, and the balanced error is $36.4^\circ$ against $35.7^\circ$ raw on the saved per-neuron predictions of held-out-animal split 0, which average 32 population draws per neuron (Table~\ref{tab:allen} reports the population-level error of the same run, $35.2^\circ$). The per-bin errors show, on both datasets, that the readout is cardinal: neurons preferring $0^\circ$ or $90^\circ$ are decoded to $11^\circ$ to $12^\circ$ on the twin and to $25^\circ$ and $34^\circ$ on Allen, oblique ones to $37^\circ$ to $39^\circ$ on the twin and at chance on Allen, and 63 to 72\% of all predictions fall on the two cardinal axes (Fig.~\ref{fig:bins}). The relations say how close a neuron is to a cardinal axis, the anisotropy of Section~\ref{sec:frame} seen from the decoder's side, and much less about which oblique it prefers.

\begin{table}[htbp]
  \centering\footnotesize
  \caption{Controls, mean angular error on held-out neurons (chance $45^\circ$), 17M decoder on 512-neuron populations unless stated; single runs unless a seed count is given. Where two numbers are given, the second is the error after the best global rotation or reflection (rows marked \emph{aligned}) or averaged over eight equally weighted true-orientation bins (rows marked \emph{balanced}).}
  \label{tab:controls}
  \begin{tabular}{llc}
    \toprule
    control & substrate & error \\
    \midrule
    standard decoder (3 seeds) & twin & $20.2^\circ$ \\
    row statistics only, no relational term & twin & $37.7^\circ$ \\
    Gram row only, no row statistics & twin / in vivo & $20.3^\circ$ / $26.5^\circ$ \\
    row statistics + Gram as attention bias (3 seeds) & twin & $19.9^\circ \pm 0.3^\circ$ \\
    disjoint stimulus bins, training vs test (3 seeds) & twin & $20.7^\circ \pm 0.4^\circ$ \\
    standard decoder (3 seeds) & in vivo & $25.4^\circ$ \\
    disjoint stimulus bins (60 per side; 3 seeds) & in vivo & $33.8^\circ \pm 0.3^\circ$ \\
    same 60 bins on both sides & in vivo & $25.3^\circ$ \\
    neuron split 1 / 2 & in vivo & $25.4^\circ$ / $24.6^\circ$ \\
    standard decoder, raw / balanced & twin & $20.3^\circ$ / $24.5^\circ$ \\
    orientation-balanced loss, raw / balanced & twin & $21.4^\circ$ / $25.3^\circ$ \\
    standard decoder, raw / balanced & in vivo & $24.6^\circ$ / $27.9^\circ$ \\
    orientation-balanced loss, raw / balanced & in vivo & $25.8^\circ$ / $29.1^\circ$ \\
    class-balanced populations (64 per class, 2 seeds), raw / aligned & twin & $22.0^\circ$ / $21.9^\circ$ \\
    class-balanced populations, raw / aligned, 2 seeds & in vivo & $26.6^\circ$ / $26.4^\circ$ \\
    trained on V1, scored on V1 (128), 3 seeds & twin & $22.4^\circ \pm 0.5^\circ$ \\
    trained on V1, scored on RL (512), raw / aligned, 3 seeds & twin & 26, 43, 46 / 25, 41, 39 \\
    trained on V1, scored on AL (128), raw / aligned, 3 seeds & twin & 58, 33, 32 / 32, 32, 32 \\
    same, class-balanced training (2 seeds), V1 / RL / AL, raw & twin & $31^\circ$ / $29^\circ$ / $45^\circ$ \\
    \quad aligned & twin & $31^\circ$ / $29^\circ$ / $36^\circ$ \\
    within-scan populations of 128, 3 seeds & twin / in vivo & $32.0 \pm 1.1^\circ$ / $31.1 \pm 1.2^\circ$ \\
    one training scan (140 neurons), populations of 128 & twin / in vivo & $38^\circ$--$44^\circ$ / $37^\circ$--$39^\circ$ \\
    mixed-scan populations of 128, 3 seeds & twin / in vivo & $22.3 \pm 0.4^\circ$ / $25.8 \pm 0.2^\circ$ \\
    \midrule
    Allen, mixed populations of 128, pool 280 / 560 / 6,706 cells & 2.2M & $40.5^\circ$ / $40.7^\circ$ / $36.0^\circ$ \\
    \bottomrule
  \end{tabular}
\end{table}

\paragraph{A subtraction ablation that leaks labels.} To remove the class-level anisotropy from the input, we estimated the class-Gram residual (8 or 24 classes) on the labelled training half and subtracted the class-pair offset from every Gram entry, using each neuron's true class, at training and test. The twin error was unchanged ($20.7^\circ$ raw and aligned at 8 classes) and improved at 24 classes ($17.7^\circ$); the control that subtracts the whole class structure, which should destroy decoding, improved it further ($17.2^\circ$ raw, $15.6^\circ$ aligned). An offset keyed to the true class is a label written into the Gram, through its jumps at class boundaries and its estimation noise, and the decoder reads it. No result of this ablation is used.

\paragraph{Synthetic populations.} 6,000 model neurons with von Mises tuning ($\kappa = 2$), preferred orientations either uniform or with half the neurons within about $12^\circ$ of a cardinal axis (count), tuning either isotropic or sharpened to $\kappa = 4$ at the cardinals (sharp), and a 120-frame stimulus that either covers orientation evenly or has half its frames near a cardinal axis (stim); responses with Gaussian noise; the 2.2M decoder on 512-neuron populations, epoch selected on the raw error as in every other run, five data and decoder seeds (Table~\ref{tab:syn}). With every switch off the class-Gram is circulant to within noise (neighbour-correlation range 0.01 against 0.04 to 0.14 with a switch on), and the decoder recovers the ordering but no frame. Sharper cardinal tuning and a cardinal excess of neurons fix the frame in at most one seed of five, alone, together, or at any strength ($\kappa$ up to 32; gain up to 5 times; two seeds each, all above $42^\circ$ raw); stimulus anisotropy fixes it in most seeds. The anisotropy the cortical data show (Section~\ref{sec:frame}) is modelled as a shared stimulus-driven component among like-tuned neurons: a random component per frame on 36 orientation nodes, smoothed over the circle with a von Mises kernel ($\kappa = 4$, half-width about $30^\circ$) so that only neurons with nearby preferences share it, enters each neuron's response with a weight peaked at $90^\circ$, with uniform counts, isotropic tuning and an even stimulus. Its class-Gram residual has the one-peaked form of the cortical one and correlates 0.39 to 0.48 with the in-vivo V1 residual; its size is set by the weight and reported as the residual's fraction of class-Gram variance (V1: 0.16 on the twin, 0.30 in vivo) together with the anisotropy factor (1.67, 1.99 and 1.92 at the three sizes). A global field over orientation with three Fourier modes also fixes the frame (four of five seeds at factor 1.32 and residual fraction 0.01, five of five at 1.68 and 0.06 and at 1.75 and 0.07), but its residual correlates negatively with the cortical one ($-0.38$), so the local field is the model reported. The spread across seeds is large in every cell where the frame is not fixed, because selection on the raw error then has no signal to select on, and the cell means are therefore reported with their range. The frame-fixed counts are unchanged with the raw-error threshold at $25^\circ$ or $35^\circ$, except that the count cell drops to 0 of 5 at $25^\circ$. One run (all three switches, seed 0) never left its initial loss and recovered neither ordering nor frame; it is included in the mean.

\begin{table}[htbp]
  \centering\footnotesize
  \caption{Synthetic populations, five seeds: raw error / error after the best global rotation or reflection, mean over seeds with the range in brackets, and the number of seeds in which the frame was fixed (raw error below $30^\circ$ and within $5^\circ$ of the aligned error). Bottom block: a shared stimulus-driven component among like-tuned neurons (a local random field over preferred orientation, weight peaked at $90^\circ$), at three sizes of the class-Gram residual it produces (V1: 0.16 twin, 0.30 in vivo). A gap between raw and aligned error means structure without frame.}
  \label{tab:syn}
  \begin{tabular}{lccc}
    \toprule
    anisotropy switched on & raw & aligned & frame fixed \\
    \midrule
    none (circulant) & $53.7^\circ$ [44--63] & $30.0^\circ$ [23--42] & 0 / 5 \\
    stimulus & $23.5^\circ$ [4--54] & $13.9^\circ$ [4--19] & 3 / 5 \\
    sharper cardinal tuning & $51.4^\circ$ [39--65] & $28.6^\circ$ [15--42] & 0 / 5 \\
    cardinal excess of neurons (count) & $37.8^\circ$ [27--45] & $35.1^\circ$ [27--42] & 1 / 5 \\
    sharper tuning + stimulus & $12.2^\circ$ [3--38] & $6.3^\circ$ [3--10] & 4 / 5 \\
    count + stimulus & $25.0^\circ$ [4--47] & $20.7^\circ$ [4--43] & 3 / 5 \\
    count + sharper tuning & $42.8^\circ$ [29--58] & $35.2^\circ$ [27--41] & 1 / 5 \\
    all three & $18.4^\circ$ [3--46] & $17.9^\circ$ [3--44] & 4 / 5 \\
    \midrule
    shared component among like-tuned neurons, residual 0.05 & $3.3^\circ$ [3--4] & $3.2^\circ$ [3--4] & 5 / 5 \\
    \quad residual 0.15 (twin V1: 0.16) & $2.4^\circ$ [2--3] & $2.4^\circ$ [2--3] & 5 / 5 \\
    \quad residual 0.23 (in vivo V1: 0.30) & $2.1^\circ$ [2--2] & $2.0^\circ$ [2--2] & 5 / 5 \\
    \bottomrule
  \end{tabular}
\end{table}

\section{Receptive-field decomposition per split}\label{app:rfsplits}
Per held-out-mouse split, for the two Allen regimes of Fig.~\ref{fig:rf}: $\Rsq$ on absolute screen coordinates, the correlation between predicted and true per-mouse mean position over the 9 test mice ($x$, $y$), and the correlation between predicted and true position relative to the mouse mean over the cells ($x$, $y$).

\begin{table}[htbp]
  \centering\small
  \caption{Allen receptive fields per split.}
  \begin{tabular}{llccc}
    \toprule
    populations & split & absolute $\Rsq$ & per-mouse mean $r$ ($x$, $y$) & within-mouse $r$ ($x$, $y$) \\
    \midrule
    single-animal & 0 & 0.29 & 0.47, 0.74 & 0.07, 0.11 \\
    single-animal & 1 & $-0.02$ & 0.48, 0.40 & 0.06, 0.08 \\
    single-animal & 2 & $-0.45$ & 0.19, 0.25 & 0.11, $-0.01$ \\
    mixed & 0 & 0.22 & 0.70, 0.81 & 0.00, 0.20 \\
    mixed & 1 & 0.08 & 0.23, 0.80 & 0.02, 0.08 \\
    mixed & 2 & $-0.25$ & 0.46, 0.75 & $-0.06$, 0.04 \\
    \bottomrule
  \end{tabular}
\end{table}

\clearpage

\end{document}